\documentclass{plateau}

\setcopyright{cc}
\copyrightyear{2025}

\plateauConference[PLATEAU '25]{15}{February 17--18, 2025}{Boston, MA}
\plateauDOI{10.1184/R1/29086778}

\begin{document} 

\title{The Teacher's Dilemma: Balancing Trade-Offs in Programming Education for Emergent Bilingual Students}

\author{Emma R. Dodoo}
\orcid{0000-0002-7957-6189}
\affiliation{%
  \institution{University of Michigan}
  \city{Ann Arbor}
  \state{MI}
  \country{USA}
}

\author{Tamara Nelson-Fromm}
\orcid{0000-0001-8197-8366}
\affiliation{%
  \institution{University of Michigan}
  \city{Ann Arbor}
  \state{MI}
  \country{USA}
}

\author{Mark Guzdial}
\orcid{0000-0003-4427-9763}
\affiliation{%
  \institution{University of Michigan}
  \city{Ann Arbor}
  \state{MI}
  \country{USA}
}

\renewcommand{\shortauthors}{Dodoo et al.}

\begin{abstract}
\textit{K-12 computing teachers must navigate complex trade-offs when selecting programming languages and instructional materials for classrooms with emergent bilingual students. While they aim to foster an inclusive learning environment by addressing language barriers that impact student engagement, they must also align with K-12 computer science curricular guidelines and prepare students for industry-standard programming tools. Because programming languages predominantly use English keywords and most instructional materials are written in English, these linguistic barriers introduce cognitive load and accessibility challenges. This paper examines teachers' decisions in balancing these competing priorities, highlighting the tensions between accessibility, curriculum alignment, and workforce preparation. The findings shed light on how our teacher participants negotiate these trade-offs and what factors influence their selection of programming tools to best support EB students while meeting broader educational and professional goals.}
\end{abstract}



\maketitle

\section{Introduction}
It is well established that English has long been the lingua franca, or \emph{``common language''}, of STEM education, in which computing is situated \citep{rajprasit_english_2015, hamel2007dominance, riemer2002english}. Programming languages (PLs) are also fundamentally grounded in English, from keywords and syntax to instructional documentation. This English-centric foundation of PLs presents a barrier for \emph{novice} programming learners whose first language is \textbf{not} English, referred to in this work as \textit{emergent bilingual} students \citep{raj_impact_2019}. Another barrier in computing classes, particularly for the EB student, is the medium of instruction in which the class is conducted; in the United States, it is English \cite{lei_english_2022, guo_non-native_2018}. 

Emergent bilingual (EB) students in the US are acquiring English despite already communicating in a primary language other than English \citep{garcia_emergent_2009}. Although EB students are the fastest-growing population in K-12 public schools, comprising 11\% of the national student population, their representation in foundational computer science courses in public schools remains disproportionately low, with only 7\%  currently enrolled \citep{barr_bringing_2011, stateofCS_2024}. This gap highlights ongoing barriers to access and participation for EB students in computing classes.

For EB students, programming classes introduce a dual challenge: learning both the \textit{spoken} English language and the \textit{technical} English of PLs to use them in classrooms, imposing a significant cognitive load \citep{soosai_raj_does_2018}. Learners must process programming concepts while simultaneously navigating linguistic complexities. In the past decade, programming instruction has expanded in K-12 settings, with some states making computer science a high school (14-18 years old) graduation requirement. As programming becomes more widely available, teachers report challenges in teaching programming, particularly in classrooms with EB students. For instance, computing teachers cite struggles with their self-efficacy in programming, which affects their confidence in adopting and integrating new pedagogies and programming tools \citep{zhou_high_2020, llosa_impact_2016}. Teachers with low self-efficacy may struggle to provide the appropriate scaffolding and resources to benefit EB students and allow them to participate effectively. Therefore, understanding the considerations teachers make to meet programming standards while supporting EB students is crucial.

To make programming more accessible, teachers must navigate trade-offs when selecting PLs and instructional strategies that align with their comfort level, academic standards, and students' linguistic needs. Accessibility in this context refers to \emph{how well EB students can understand programming concepts, languages, and tools, regardless of their English proficiency or prior CS exposure} \citep{shrestha_here_2022, becker_parlez_2019}. 

Contributing to the greater bilingual computing education research, we explore how our teacher-participants make decisions about and, thus, prioritize which PL tools are selected for general use in their programming classrooms. The research question to examine the balancing trade-offs teachers consider in this paper is: \emph{\textbf{What trade-off factors shape K-12 computing teachers' decisions when selecting and implementing programming tools to support emergent bilingual students?}}
We identify four (4) thematic findings in response to the research question: to understand the instructional goals and the tensions between them. Broadly, the trade-offs align with workforce preparation, without overwhelming learners, by balancing cost and engagement for exploration, minimizing extraneous load through accessible representations, and scaffolding learning with familiar and intuitive syntax.

\section{Prior and Related Work}
The literature on bilingual computing education continues to expand within and beyond the United States. In the US, much of the research has focused on student outcomes, with a dominant emphasis on elementary and middle school students \citep{vogel_role_2019, jacob_examining_2022, jacob_elementary_2024}. Internationally, studies in bilingual computing education have also centered on student outcomes, primarily at the undergraduate level \citep{pal_framework_2016, pal_framework_2016}. This focus leaves a gap in research on high school students and teachers, limiting our understanding of how bilingual computing education is implemented at this critical stage.

Additionally, little research examines the programming tools used in bilingual computing classrooms despite their potential to introduce cognitive strain and impact student learning. In this section, we discuss cognitive load theory, teachers' perceptions of EB students' challenges, literature on the medium of instruction, and the role of non-English programming languages in computing education.


    \subsection{Teacher Perceptions and Challenges in EB-Inclusive Programming Classrooms} 
    Programming requires learners to engage in cognitively demanding tasks simultaneously. The tasks include problem-solving, program design, debugging, and mastering the syntax of the programming language \cite{denny_understanding_2011}. In Dodoo et al.'s (\citeyear{dodoo_teaching_2025}) study, they discuss how strategies explicitly used to support EB students can also benefit non-EB students, which aligns with the concept of the ``curb-cut effect'' as defined within Universal Design \citep{johnson2003creating}.

    Prior research also explains that K-12 CS teachers frequently describe their computing self-efficacy as challenging to adopt and integrating new tools into their classroom \cite{zhou_high_2020, llosa_impact_2016}. Teachers with low self-efficacy may struggle to offer the scaffolding and resources necessary to support EB students in succeeding in programming. On the contrary, teachers with higher computing and pedagogical self-efficacy are better equipped to design engaging activities, provide targeted feedback, and create inclusive learning environments. So, it is important to understand the considerations that teachers must, or would be more willing to make, the changes that could better support EB students.

    \subsection{Teaching Computing in a Non-English Format}
    The language of instruction influences EB students' progress, motivation, and engagement in programming education. In Botswana, Tshukudu et al. (\citeyear{tshukudu_bilingual_2024}) conducted a national programming outreach initiative, dividing students from sixteen (16) Senior Secondary Schools into two groups: one programming in English and another using a bilingual approach (Setswana and English). Their findings revealed that students in the bilingual group showed statistically significant improvements in their it programming comfort level than the English-only group, suggesting that using a familiar language can create a psychologically safer \citep{pedersen_authenticity_2022} and more supportive learning environment for the bilingual students. 
    
    In another study, Raj et al. (\citeyear{raj_impact_2019}) investigated the impact of incorporating Tamil in an undergraduate data structures course that used C++ as the programming language. While no statistically significant differences in student learning outcomes were observed between the  English-only and bilingual (Tamil and English) groups, Tamil-speaking students in the bilingual group demonstrated higher engagement and comfort levels in classroom interactions and discussions about programming concepts (e.g., asked more questions and asked questions of higher complexity). Building on this, Molina et al. (\citeyear{molina_effects_2023}) replicated the above study to examine what influence language has among Spanish-speaking students in the US. They found that high school students from the bilingual group (Spanish and English) were more engaged and comfortable asking questions. However, no statistically significant differences in student learning were seen among the English-only and bilingual groups. However, this further suggests that while bilingual instruction may not directly influence measurable learning outcomes, it can improve classroom dynamics and student participation.
    
    In his dissertation, Pal (\citeyear{pal_framework_2016}) explored best practices for teaching programming to native Hindi speakers. His work revealed that while these students were motivated to learn in English, they achieved the best outcomes when instruction was provided in their native language or the medium of instruction they had experienced during K-12 education. Pal recommended strategies such as providing recorded lectures in English, allowing students to revisit the material content at their own pace, looking up unfamiliar terms, and gradually making sense of programming concepts. This approach acknowledges the cognitive and emotional advantages of incorporating native language support in programming education.
    
   These studies emphasize that integrating native languages into PLs can improve student engagement. 

    \subsection{Non-English Programming Languages}
    Beyond instructional challenges, traditional text-based programming languages (TBPLs) present language and accessibility barriers for EB students due to the English-centric nature of syntax and keywords and their limited support for exploratory learning. Non-English or bilingual PLs demonstrate how language barriers can be learned and how programming tools can be designed to encourage exploration.

    Block-based programming languages such as Scratch, Snap, Code.org's Blockly, and MIT AppInventor offer visual approaches that reduce syntax complexity and provide immediate feedback when users drag and drop blocks to construct their programs \cite{liao_scratch_2023, baulearnable}. Some BBPLs have been localized to support other non-English languages (e.g., Scratch and Snap!), which enables EB students to code in their native language or seamlessldo y translanguage between English and their native tongue \cite{vogel_role_2019, sarasa_use_2019} by toggling the `language' button.

    Task-specific programming languages ~\cite{guzdial_task_1997, guzdial_putting_2023}, like Pixel Equations \cite{dodoo_designing_2022}, have been designed to reduce language barriers in computing classrooms by allowing students to type certain words in Spanish (e.g. \textit{rojo}, \textit{verde}, and \textit{azul}) and toggle between Spanish and English instruction interfaces. It was developed for an 11th-grade engineering class and is also being used in an undergraduate \emph{application of computing for creative expression at a large R1 institution} \cite{guzdial_a_2024}. Another example of a teaspoon language is Charla-bots~\cite{guzdial_putting_2023} and Data Visualization for Learning~\cite{shreiner_the_2022, naimipour_guided_2021}.

    Localized programming languages \cite{localizedPLs} have also been designed to support non-English learners who want to program. These PLs are not nearly as widespread or represented in bilingual computing education literat; however,but they exist. Some localized PLs that have been translated include Python to Chinese~\cite{chinesePython}. At the same time, French has a localized version of BASIC called LSE (Langage Symbolique d'Enseignement)~\cite{frenchLSE}, and Pascal, an older but well-known PL, has a Portuguese version called VisuAlg~\cite{deSouza_revista_2009}. Non-English programming languages, which are foundationally designed to be in a language other than English, such as Easy-PL (Chinese), Linotte (French), Dolittle (Japanese), and PSeInt (Spanish) have served regions with limited access to English-based computing resources, fostering more equitable opportunities for students in countries across Latin America (e.g., Argentina, Bolivia, Chile, Colombia, Ecuador, Paraguay, Peru, Uruguay and Venezuela) \cite{castrillon_analysis_2022}.


\section{Methodology}

\subsection{Recruitment}
We recruited participants through Twitter, the first author's website, K-12 computing education mailing lists, and by outreach to local school districts with connections to the previous study team. This is a form of purposive sampling (i.e., sampling that selects participants based on specific criteria that align with a study's objectives) \citep{tongco_purposive_2007}. There were two main eligibility criteria for teachers to be considered as participants. The first was that the teacher participants must (1) be teaching a computing or programming class and (2) have EB students in their programming classrooms. We initially had 27 teacher participants who expressed interest in the study. However, a significant number (19) were excluded because they could not provide a valid school email address\footnote{The decision was made following incidents where some original participants joined Zoom sessions with unprofessional or suspicious postures, which caused discomfort among the interviewers.} Further investigation revealed that in some cases, the email addresses provided were personal rather than professional or were associated with outdated school systems. While some potential participants seemed genuinely interested in the research, a verifiable school email address was necessary to confirm their employment as K-12 computing teachers within the participating school systems and ensure reliable communication throughout the research process.

    \subsection{Design Probes and Data Collection}
    Data were collected through semi-structured interviews \cite{merriam2015qualitative}, each lasting approximately 65 minutes. The interviews aimed to explore educators' perceptions of various PL by presenting them with identical pieces of code in different programming environments. We sought to understand participants' preferences, challenges, and perceived potential of using these tools in their classroom.

    The participants were introduced to PLs from different categories, namely:
    \begin{itemize}
        \item \textit{Block-based programming languages (BBPLs)} because they offer a visual and intuitive interface, minimizing the cognitive burden of syntax. The participants were shown short example programs in Scratch, presented in both English (see Fig. \ref{fig:bbpl-en}) and Spanish  (see Fig. \ref{fig:bbpl-es}).
        
        \item \textit{Text-based Programming Languages (TBPLs)}, as these require specific syntax, symbols, and keywords. These languages offer greater flexibility but a steeper learning curve, particularly for novice programmers. Participants were presented with examples of C++ and Python.
    
        \item \textit{Text-specific Programming Languages (TSPLs)} because they are designed for domain-specific tasks and have tailored interfaces. For example, Pixel Equations allows users to create custom image filters using equations to describe sections of an image, similar to graphing \cite{dodoo_designing_2022}. Colors can be defined through equations, and conditional statements allow users to test and select pixels based on their values. Participants reviewed examples from Pixel Equations (shown in both English and Spanish Fig. \ref{fig:tbpl-pixeq-en} and Fig. \ref{fig:tbpl-pixeq-es}) and Charla-bots (Fig. \ref{fig:tbpl-charla}).
     
        \item \textit{Gradual programming languages (GPLs)} as these introduce programming concepts incrementally. Hedy \cite{hermans_hedy_2020}, which was presented in both  English and Spanish Fig. \ref{fig:gpl-en} and Fig. \ref{fig:gpl-es}), served as the example GPL.
        \begin{figure}[h]
    \centering
    \begin{minipage}{0.45\textwidth}
        \centering
        \includegraphics[width=\linewidth]{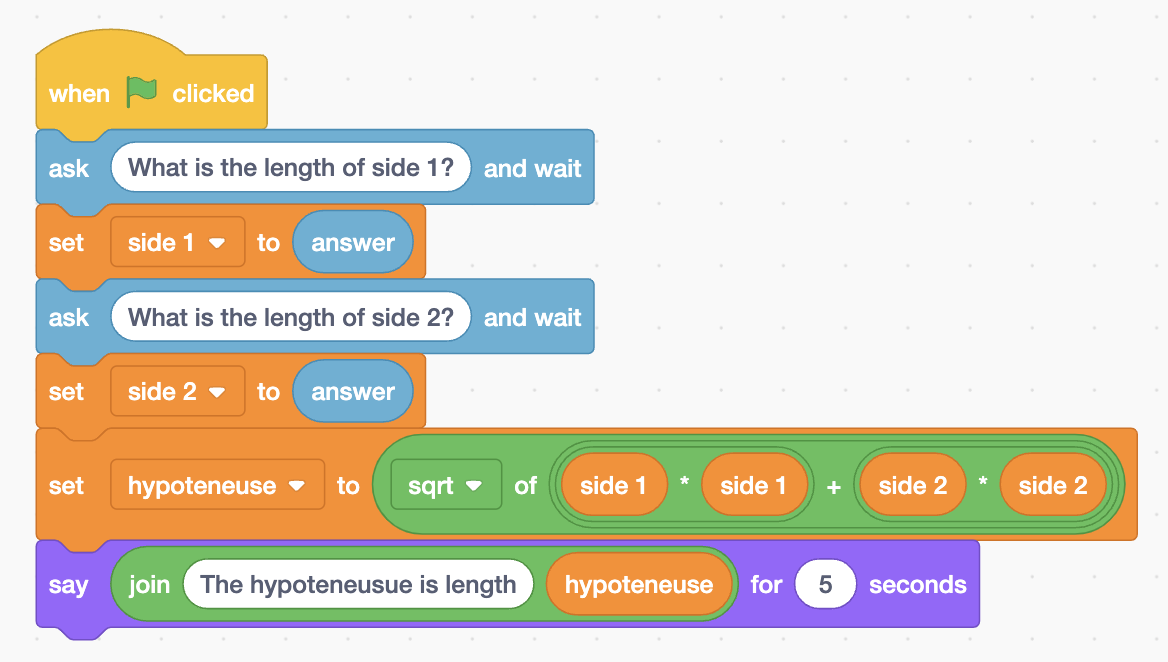}
        \vspace{-20pt}
        \caption{Scratch (English)}
        \label{fig:bbpl-en}
    \end{minipage}
    \begin{minipage}{0.45\textwidth}
        \centering
        \includegraphics[width=\linewidth]{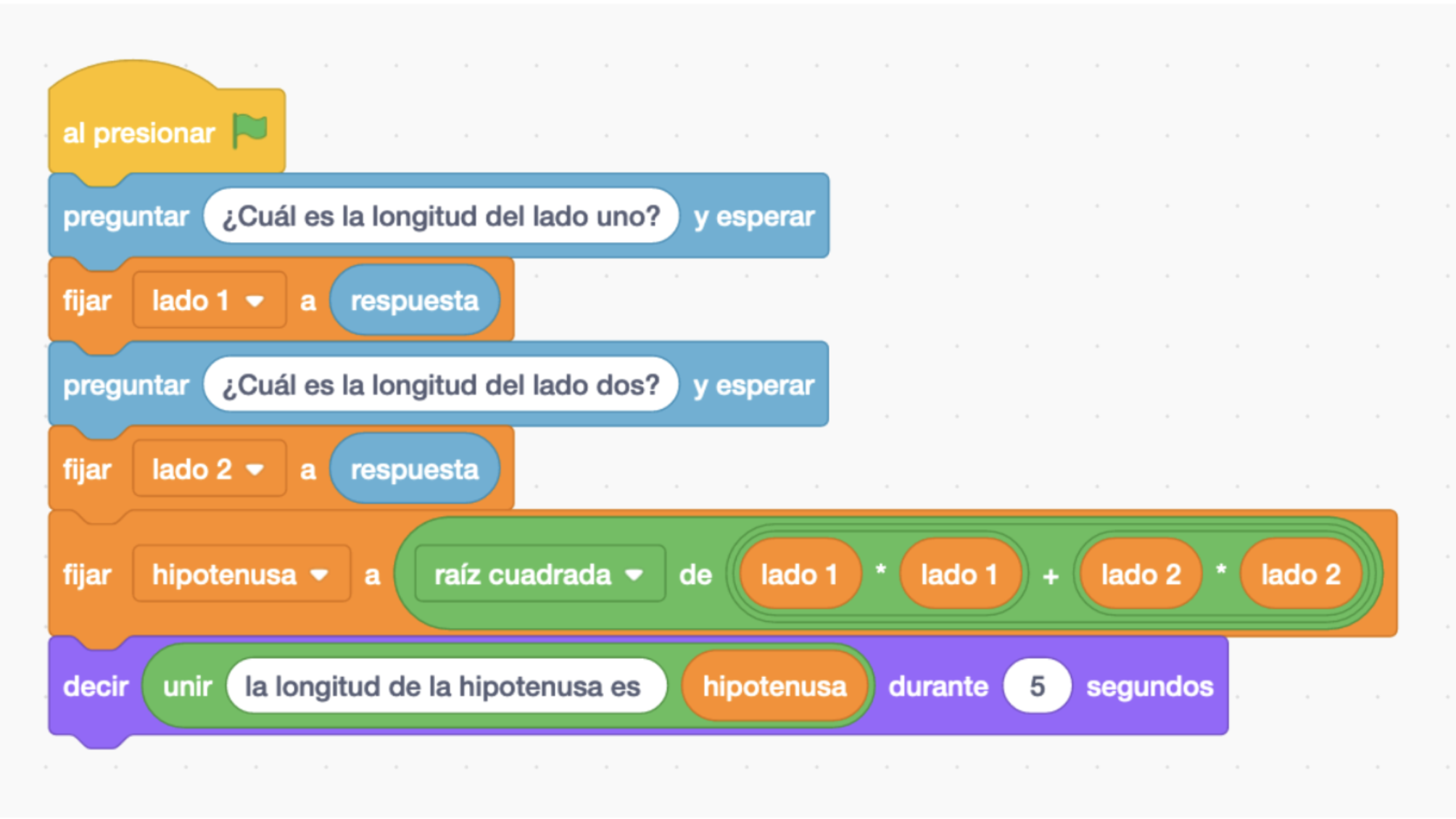}
        \vspace{-20pt}
        \caption{Scratch (Spanish)}
        \label{fig:bbpl-es}
    \end{minipage}
    \begin{minipage}{0.32\textwidth}
        \centering
        \includegraphics[width=\linewidth]{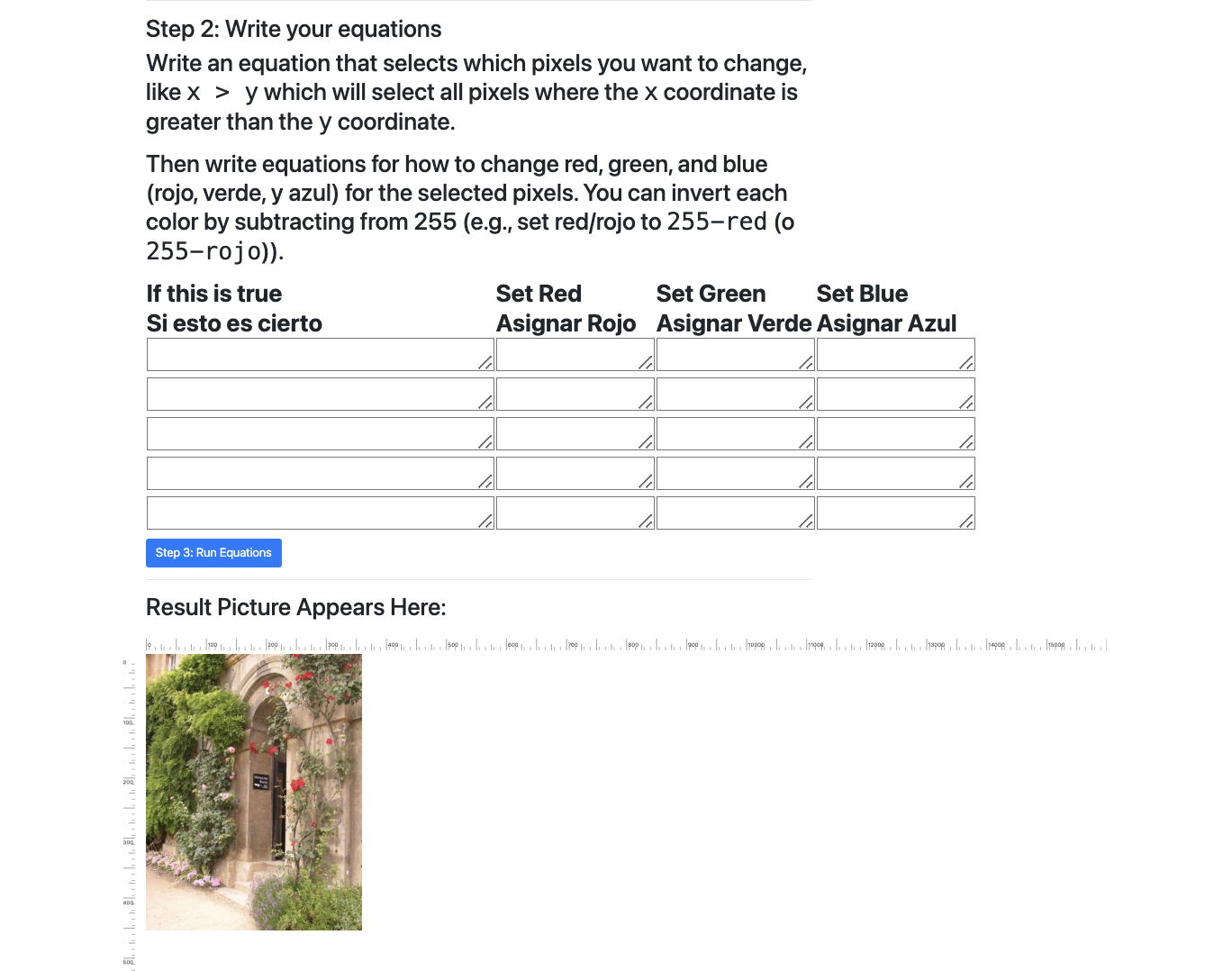}
        \vspace{-20pt}
        \caption{Pixel Equations (English)}
        \label{fig:tbpl-pixeq-en}
    \end{minipage}
    \hfill
    \begin{minipage}{0.32\textwidth}
        \centering
        \includegraphics[width=\linewidth]{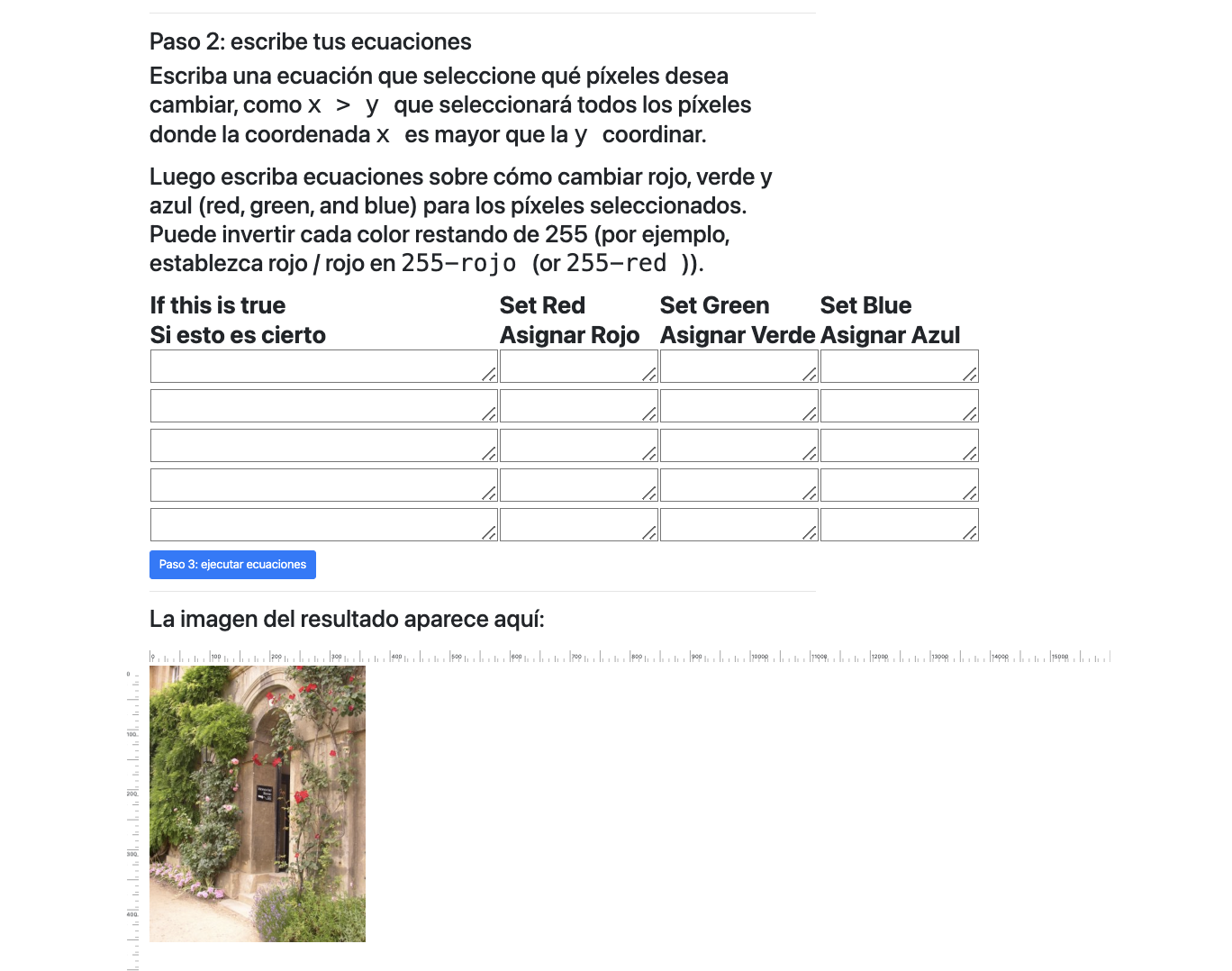}
        \vspace{-20pt}
        \caption{Pixel Equations (Spanish)}
        \label{fig:tbpl-pixeq-es}
    \end{minipage}
    \hfill
    \begin{minipage}{0.32\textwidth}
        \centering
        \includegraphics[width=\linewidth]{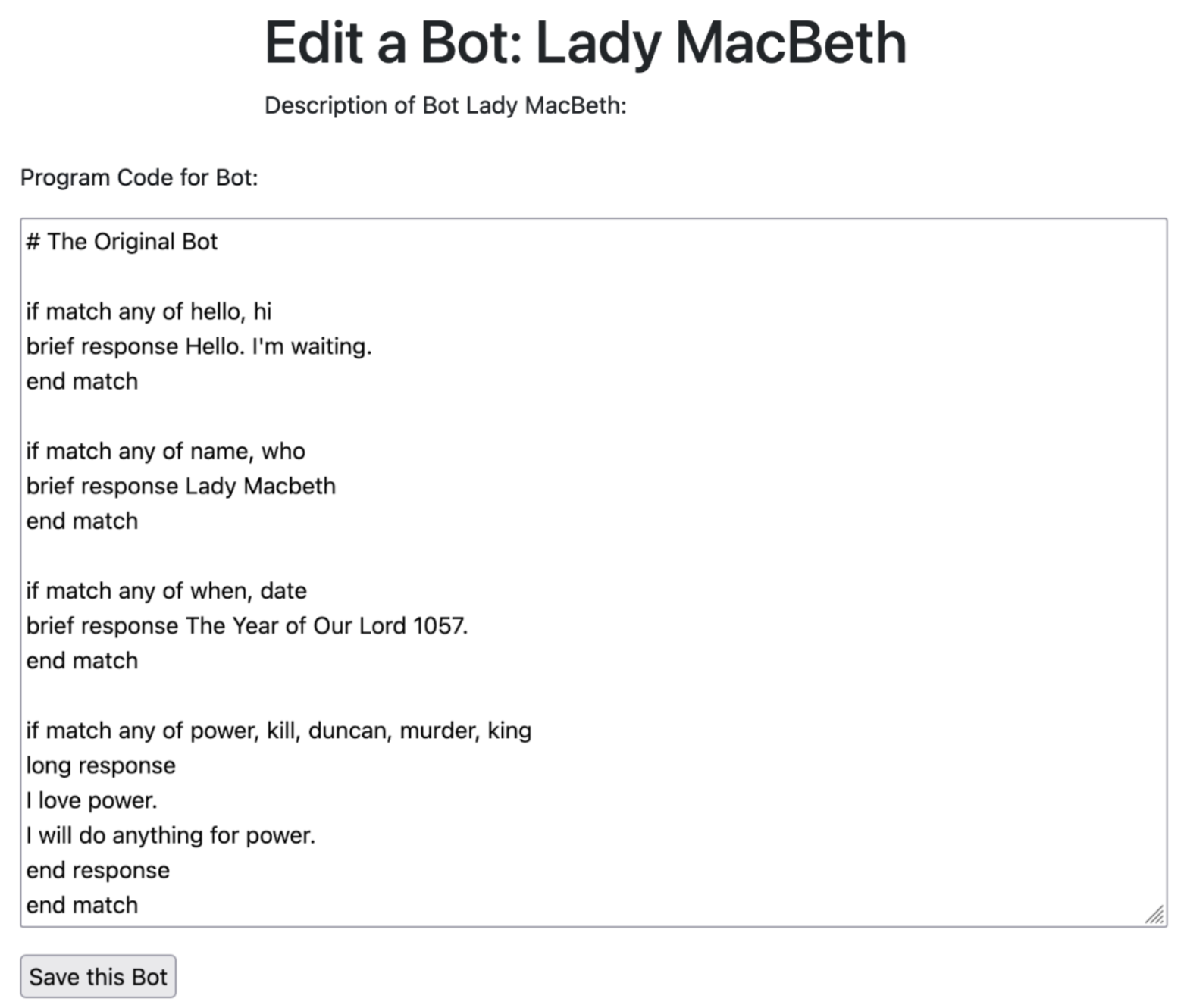}
        \vspace{-20pt}
        \caption{Charla-bots}
        \label{fig:tbpl-charla}
    \end{minipage}
    \begin{minipage}{0.4\textwidth}
        \centering
        \includegraphics[width=\linewidth]{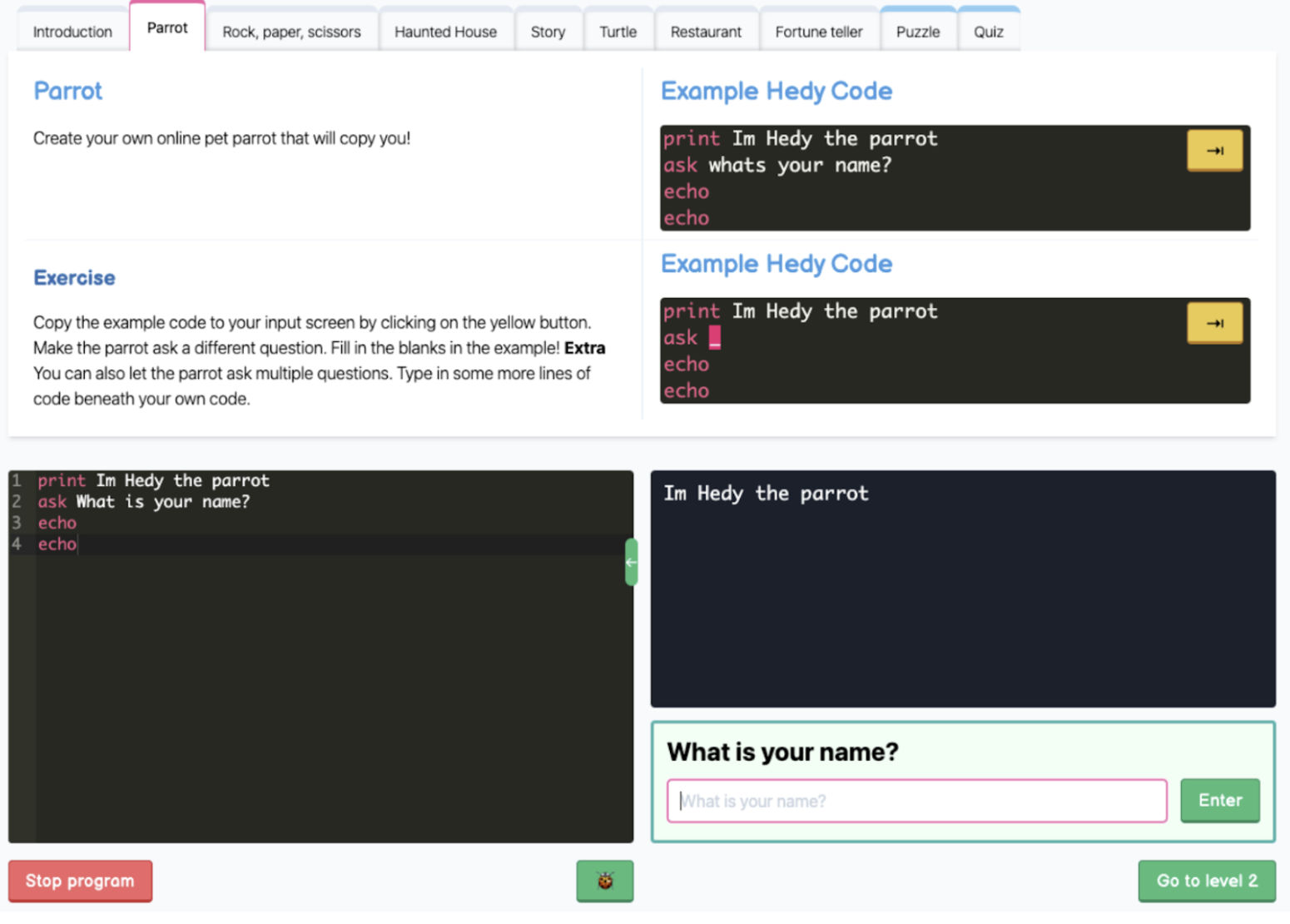}
        \vspace{-20pt}
        \caption{Hedy (English)}
        \label{fig:gpl-en}
    \end{minipage}
    \begin{minipage}{0.4\textwidth}
        \centering
        \includegraphics[width=\linewidth]{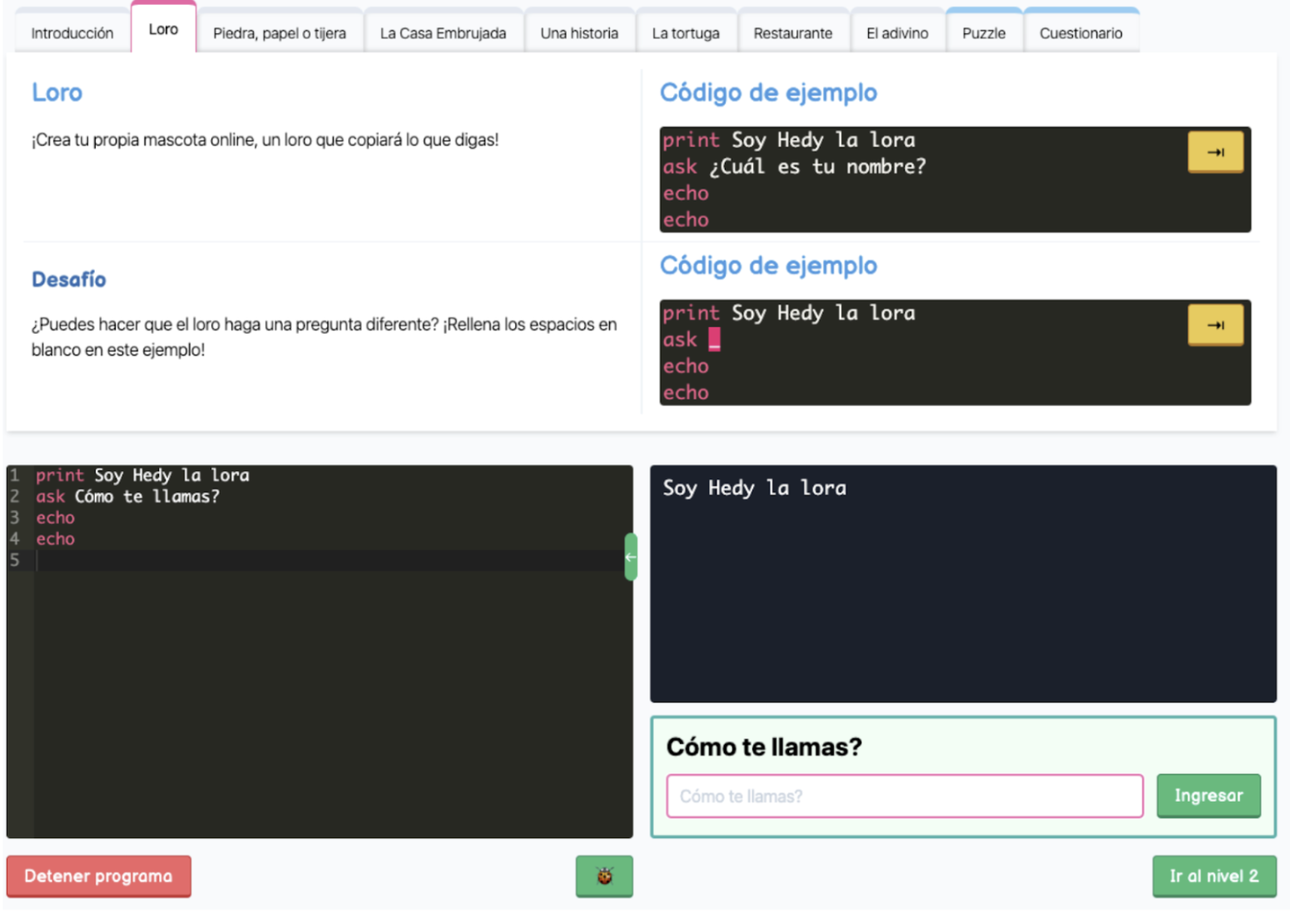}
        \vspace{-20pt}
        \caption{Hedy (Spanish)}
        \label{fig:gpl-es}
    \end{minipage}
\end{figure}
    \end{itemize}
        
    Each PL type was introduced individually. The PL was viewed on a shared screen as interviews were conducted over Zoom. Participants were asked the same questions for each PL, irrespective of the language in which the PL was presented. Example questions included: ``What important characteristics or features do you look for when choosing, using, or implementing a programming language in your classroom?'' and ``What would you like to see in PLs that support CS learning goals?''.

    \subsection{Data Analysis and Validation}
    Recordings of the interviews were transcribed using Rev's transcription software\footnote{\url{www.rev.com}}, then de-identified and reviewed for accuracy. The familiarization process involved listening to the recordings and repeatedly reading the transcripts to ensure deep engagement with the data. This iterative process facilitated an initial understanding of emerging patterns and insights.  

    The interview transcripts were analyzed using reflexive thematic analysis \citep{braun_reflecting_2019}. The analysis began with organizing quotes and notes from the transcripts using a digital sticky-note format on Miro\footnote{\url{www.miro.com}}. An affinity diagramming approach \citep{kawakita1991original} was employed to cluster related excerpts and notes into initial thematic categories. Given the first author's (ERD) background in human-centered design, the data was approached through a lens informed by the values and traditions of that field. These broad categories were iteratively refined by the second author, merging similar groups, and then we reorganized them into more focused clusters. Supporting examples were identified within the data to substantiate the emerging themes. Multiple discussions among the research team helped review and refine these themes, ensuring they accurately reflected the data. This iterative refinement process resulted in a set of prototype themes further developed into the finalized trade-offs presented in this study.  

    To strengthen the validity of the findings, the first author conducted follow-up conversations via Zoom with each willing participant. This process, known as member-checking \citep{birt_member_2016}, allowed teacher participants to review and confirm the identified themes identified by the authors. In the conversations, the teacher-participants provided additional examples and clarified statements, further strengthening the reliability of the results.
    
    \subsection{Participants}
    The study included K-12 computing teachers from different regions and educational settings across the United States. The participant group was evenly split by gender and included individuals who self-identified as White (4), Latiné \footnote{Latiné is a recent term used to describe Spanish-speaking individuals.} (4), and Black (4). This term acknowledges historically problematic epistemologies. Latiné is also easier to pronounce in Spanish since it is phonologically attuned \citep{villanueva_alarcon_latine_2022}. (3), and biracial (1). Their experience with PLs varied: half primarily used block-based languages, while the other half had experience with block-based and text-based languages. Geographically, five (5) teachers were based in Texas (covering areas along the west border, southeast, and southwest), and three (3) were in Michigan (central and southeast regions). All participating teachers reported that most of their EB students were Spanish speakers, with two (2) teachers mentioning having students who also spoke French or Korean. The predominance of Spanish-speaking EB students aligns with Spanish being the most widely spoken non-English language in the US \citep{dietrich_nearly_2022}. Additional details about participants' teaching experience, language skills, certifications, and grade levels are provided in Table \ref{tab:demographics}.

\begin{table*}
\centering
\caption{Description of Teacher Participants by Pseudonym, (b=beginner; i=intermediate; and f=fluent)}
\label{tab:demographics}
\resizebox{\textwidth}{!}{%
\begin{tabular}{lllllll}
\toprule
    \textbf{Pseudo.} &
    \textbf{Total Years Teaching} & 
    \textbf{Race/Ethnicity} &
    \textbf{Grade Level} &
    \textbf{Familiarity in Teaching} & 
    \textbf{Programming Confidence} & 
    \textbf{Language(s) Spoken}
\\ \midrule
Alison & \textit{10-12} years & White & K-2 & \begin{tabular}[c]{@{}l@{}}{Robotics and }\\{Digital Citizenship}\end{tabular} & \begin{tabular}[c]{@{}l@{}}{Math, Reading Specialist, } \\{K-8 General Education}\end{tabular} & English (f) \\
        &    &      &                                  &                                  \\
Jessica & 10-12 years& White & 3-5  & \begin{tabular}[c]{@{}l@{}}{Robotics and }\\{Programming}\end{tabular} & \begin{tabular}[c]{@{}l@{}}{Technology and }\\{Computer Science}\end{tabular} & English (f); Spanish (i) \\
        &    &      &                                  &                                  \\
Tabitha & 19-21 years & White & 7-8 & \begin{tabular}[c]{@{}l@{}}{CS First and }\\{CS Discoveries}\end{tabular} & Computer Science & {English (f); Spanish (b)} \\
        &    &      &                                  &                                  \\
Clara & 19-21 years & White & 9-12 & \begin{tabular}[c]{@{}l@{}}{AP CS Prinicples }\\{and AP CS A}\end{tabular} & Computer Science & English (f); Other (b) \\
        &    &      &                                  &                                  \\
Diego & 1-3 years & Biracial & 9-12 & \begin{tabular}[c]{@{}l@{}}{IT, Networking, }\\{and Cybersecurity}\end{tabular} & Technology & English (f); Spanish (b) \\
        &    &      &                                  &                                  \\
Javier & 4-6 years & Latiné & 9-12 & CS One and AP CS A & \begin{tabular}[c]{@{}l@{}}{Math, Technology,}\\{Computer Science}\end{tabular} & English (f); {Spanish (f)} \\
        &    &      &                                  &                                  \\
Miguel & 4-6 years & Latiné & 9-12 & \begin{tabular}[c]{@{}l@{}}{AP CS Principles, }\\{Cybersecurity, and}\\{Game Design}\end{tabular} & Math, Computer Science & English (f) \\
        &    &      &                                  &                                  \\
Oscar & 7-9 years & Latiné & 9-12 & \begin{tabular}[c]{@{}l@{}}{AP CS Principles, }\\{AP CS A and Statistics}\end{tabular} & \begin{tabular}[c]{@{}l@{}}{Math, Technology,}\\{Computer Science}\end{tabular} & English (f); Spanish (f) \\
\bottomrule
\end{tabular}%
}
\end{table*}
    
\section{Thematic Findings}
Our teacher participants' insights into their experiences and preferences for PL characteristics and features. The findings are based on what the authors subjectively understand to be the important trade-offs or priorities that teacher participants must consider when selecting appropriate PL tools and environments to support EB students. \footnote{Note: To improve the readability, direct quotations were edited to remove conversational filler words.}

    \subsection{Balancing cost and engagement for exploration} \label{theme1}
    A key consideration for teachers is fostering exploration and experimentation as a means of learning, particularly through tools and activities that promote \emph{safe} engagement. As Miguel described:
    \begin{quote}
        \textit{``They need to feel safe in my class. Once they feel safe, they're more willing to code, explain, and talk. Mistakes are acceptable, and there's no judgment or mockery.''}
    \end{quote}
    Students are more likely to engage and feel comfortable volunteering in a safe space, even if they make mistakes. This sense of security also encourages students to explore programming tools more freely. While no teacher in this study explicitly described a PL as making students feel unsafe, they highlighted how the environment and instructional approach influence students' willingness to engage. 
    
    Alison explains how trial and error is a way she sees her EB students navigating a BBPL: 
    \begin{quote}
        \textit{``If my [emergent bilingual] students struggle, I see them clicking [in a trial and error manner]. So, I tell them to put something down and then click play or run to see what happens, and then you fix it from there.''}
    \end{quote}
    Alison's description illustrates how students engage with programming by \textit{playing around with} or testing different possibilities. This approach allows them to focus on programming logic rather than getting stuck on syntax or text-heavy guiding instructions. Teachers noted that tools with immediate visual feedback supported this type of trial-and-error engagement, reducing frustration. 

    Some teacher participants also described how structured time for reflection contributed to students' ability to engage effectively with programming tasks. As Clara put it:
    \begin{quote}
        \textit{``It is necessary that EB students have adequate think-time and an opportunity to write or code their thoughts. Providing EB students time to clarify thoughts is very helpful.''}
    \end{quote}
    This additional time allows students to process the content at their own pace, helping them overcome language barriers and articulate their ideas more effectively in English or their native language. This is particularly relevant when dealing with the cognitive load of translating abstract programming concepts and syntax into their native language. 
    
    These observations 
    highlight the teachers' balancing of engagement and accessibility when selecting programming tools and designing activities. While BBPLs were frequently cited as beneficial programming entry-ways for EB students, teachers also considered factors such as long-term skill development and alignment with K-12 Computer Science learning goals when choosing tools for their classrooms.

    \subsection{Aligning with workforce preparation without overwhelming learners} \label{theme2}
    While the teachers want their students to view PLs as familiar and relatable, they also prioritize career readiness. They believe that learning industry-standard programming tools, such as the TBPLs, better equips students for \emph{`real world'} professions. Early exposure to Java, Python, or C++ likely means these students will have more experience with TBPLs, better positioning them for competitive technological careers such as software development. This emphasis on workforce preparation is particularly important for EB students, many of whom are encountering programming for the first time. Oscar highlighted this in his teaching approach:
    \begin{quote}
        \textit{``For my emergent bilingual students, this is their first programming language exposure. I show them aspects of professional programmers' lives.''}
    \end{quote}

    Beyond simply introducing students to programming, some teachers focused on guiding them through the structured thought processes required in professional software development. Javier expanded on this idea, emphasizing the importance of problem-solving and structured thinking:
    \begin{quote}
        \textit{``Programming, for me, is enabling emergent bilinguals to create programs based on specifications. They must begin thinking about the process, then be able to break code into parts, and test to make sure they progress correctly, rather than building a whole program that goes awry. It's training them to think like software engineers.''}
    \end{quote}
    The teacher participants expressed a shared aspiration for their EB students to pursue careers such as becoming professional programmers, a goal they believe is reinforced by parents and academic counselors. However, they also recognized students' challenges in learning or transitioning between different programming environments.

    In particular, high school teachers stressed the need for programming tools to support students moving from BBPLs to TBPLs—or vice versa. To facilitate this transition, they highlighted the importance of scaffolding, such as integrating buttons that translate text-based code into digestible visual representations, similar to those in BBPLs. Some widely used programming languages, such as C++, have explored similar transitional features \cite{kolling_frame_2017, federici_a_2011}. As students build confidence and familiarity with programming concepts, they should have opportunities to transition to TBPLs, allowing them to tackle more complex and ‘real-world’ challenges.

    \subsection{Minimizing extraneous load through accessible representations} \label{theme3}
    Our teachers recognized that programming can be particularly challenging for first-time programmers, including EB students. To address these challenges, they discussed how they seek TBPLs that have \emph{visual representation} (e.g., such as color coding \cite{weintrop_bloks_2017}) to support further understanding of programming concepts. Visual aids that mimic the `accessibility' offered by BBPLs could perhaps connect better with abstract programming concepts and language syntax. As Jessica said:
    \begin{quote}
        \textit{``It's important for emergent bilingual students to understand coding by engaging step by step, through visual aids like color, hands-on activities, or both. They can then apply their knowledge to tools, which allows them to see their code in action.''}
    \end{quote}

    Teachers noted that providing students adequate time to interact with these tools and engage in experimentation is crucial. EB students, in particular, benefit from opportunities to work through programming challenges at their own pace, gaining confidence through iterative problem-solving. Creating a safe space for students to make mistakes is equally important; it encourages a mindset where mistakes are seen as part of the learning process, not as failures. Visuals, such as images and color, reengage students and increase engagement. Tabitha notes:
    \begin{quote}
        \textit{``Visuals are important and helpful for getting my [emergent bilinguals] students' attention and strengthening their understanding''}
    \end{quote} 

    The ability to see the immediate effects of their code in a visual format helps EB students focus on understanding programming logic rather than getting stuck on syntax or language barriers. Visual design, such as using color-coding and spacing, can also address students' reluctance to engage with dense blocks of text. Tabitha reflected on this challenge and said:
    
    \begin{quote}
        \textit{``Color coding and spacing are necessary. My [emergent bilingual] students do not read [instructional] paragraphs unless they are explicitly directed to do so.''}
    \end{quote}

    These tools lower the cognitive load associated with programming and foster engagement by making abstract concepts tangible for EB students. Teachers also highlighted how this approach aligns with broader educational strategies, such as scaffolding and differentiated instruction, which are essential for meeting EB students' cultural and linguistic needs.

    Teachers recognized that programming is challenging for first-time learners (and EB students), and they expressed ways they wanted to help students using TSPLs visualize their code similarly to BBPLs.

    They felt that having a visual representation, such as color, would make programming more accessible. For example, in the elementary grades (5-10 years old) and middle school grades (11-13 years old), teachers frequently use BBPLs because they offer a visual representation of code, making it easier for students to understand and connect with programming.

    \subsection{Scaffolding learning with familiar and intuitive syntax} \label{theme4}
    Participants emphasized that a PL holds greater value when its syntax \emph{closely} resembles or mirrors the structure and patterns of natural language (i.e., English). Teachers particularly recognized the challenge for EB students who must navigate "spoken English" and "technical" English for programming. This concern was more frequently raised by high school programming teachers, who observed students struggling with syntax-heavy languages (e.g., C++ and Java). Simplified syntax has the opportunity to reduce the cognitive load EB students experience \cite{cheung_plain_2017, guo_non-native_2018} due to code-switching between domain terms and their primary language (e.g., Spanish) \cite{dodoo_teaching_2025}. 
    Alison shared her own experience of learning a TBPL: C++. Alison said that:
    \begin{quote}
        \textit{``Memorizing C++ is difficult. I would write some code, thinking I had it correct, but then realize I had forgotten a small detail. It's like memorizing letters or words in chunks, including quotations and all those elements.''}
    \end{quote}
    
    Here, we see syntax memorization contributing to cognitive load, particularly the need to recall exact symbols and structures. Cognitive Load Theory (CLT) \citep{sweller_evidence_1991} suggests that excessive intrinsic load—such as remembering syntax details—can hinder learning, especially for novices and EB students. Teachers noted that this challenge also extended to students, making programming languages with complex syntax harder for them to adopt.
    
    However, teachers expressed a notable shift in attitudes when discussing Python, another TBPL. Miguel best explains this as
    \begin{quote}
        \textit{``Python is the language I see where my [emergent bilingual] kids say, `Okay, it makes sense. There's not so much translation. There are cognates, and it's simpler.' They really like Python code, especially from year one to year two.''}
    \end{quote}
    Here, Miguel highlights how Python's simplified syntax, use of cognates, and reduced symbolic complexity contribute to greater student engagement. Unlike Alison's struggles with C++, Miguel's observation focuses on student experiences, suggesting that teachers' attitudes toward PLs may be shaped by their learning journeys and their students’ successes or difficulties. 

    Some other teachers argued that PL designers might sidestep these difficulties by using primarily non-linguistic symbols instead of English words. However, participants also pointed out that symbolic-heavy languages present their cognitive challenges. Clara noted:
    \begin{quote} 
        \textit{``The reason I like Python is its simpler syntax. JavaScript has many extra symbols. Every letter is a symbol, and curly braces and semicolons require precise placement. Learning a new language with so many symbols is difficult.''}
    \end{quote}

    Beyond reducing cognitive burden, familiarity with syntax boosted student confidence and engagement among  EB students, allowing them to draw on their prior knowledge to connect new concepts to what they already know. 


\section{Discussion}
This study reveals the complex trade-offs the participating teachers of this study must consider when selecting PLs for EB students. The pedagogical challenge involves balancing linguistic accessibility, syntactical complexity, and academic rigor while fostering student engagement. Prior research informs researchers about how the English-centric foundation of PLs presents a significant barrier for EB students \citep{guo_non-native_2018}. To address this, teachers strive to scaffold learning, prepare students for future career opportunities, and cultivate meaningful engagement with computing. 

\textbf{Scaffolding and cognitive load management :} The themes discussed in the findings, namely, `balancing cost and engagement for exploration' and ` learning with familiar and intuitive syntax,' demonstrate the role of scaffolding in teaching programming to EB students. Aligned with cognitive load theory, syntax that resembles natural language helps lower students' \textit{extraneous} cognitive load by aligning programming logic with students' existing linguistic knowledge \cite{guo_non-native_2018}. As Pal (\citeyear{pal_framework_2016}) suggests, providing support in a familiar language can create a more supportive learning environment. Teachers perceive BBPLs as a critical entry-way for EB students, as they ease the transition into programming by making the code more intuitive and accessible \cite{weintrop_to_2015}through their use of color, immediate feedback mechanisms, and visual representations, are considered more approachable, thereby reducing the extraneous cognitive load associated with abstract syntax and error-prone text-based coding. These approaches have the opportunity to scaffold learning and support a more intuitive understanding of programming concepts, thus facilitating deeper engagement by reducing language barriers and addressing the challenges EB students face in understanding programming concepts, languages, and tools, regardless of their current proficiency in English.




\textbf{The Tension with Preparing Students for the Workforce}: While the teacher participants allude to how PLs that resemble natural language have the opportunity to reduce EB students' extraneous cognitive load, how BBPLs have prove to have an easier entry providing students with the opportunity to `explore' on their own, there is a contention in preparing students for future technological careers. As seen in the theme "aligning with workforce preparation without overwhelming learners," the high school teacher participants prioritized TBPLs to prepare EB students for jobs requiring some level of programming. The focus on TBPLs, while necessary for career preparation, if introduced too early to EB students, can increase the cognitive load for EB students due to the languages' complex syntax. As Becker (\citeyear{becker_parlez_2019}) notes, there are significant barriers to PL acquisition for non-native English speakers. It is important to note that teachers also think about preparing their high school students, not including EB students, for technological careers, and thus have to prepare them for these jobs. When given the opportunity, they also partner with industry/corporate partners. However, TBPLs, for teachers in this study, need to be used because they are the closest PLs that resemble what they will use in their vocation/career. BBPLs and TSPLs (e.g., Pixel equations) were seen to be entry way programming tools and were preferred at the earlier grade levels (K-8), while TBPLs were preferred at high school. 


\textbf{Balancing Engagement with Rigorous Preparation}: The challenge of balancing scaffolding for accessibility with preparing students for the future workforce was a central concern for our teacher participants. They aim to provide EB students with opportunities to explore and experiment, fostering confidence and engagement. Activities such as creating photo filters or using `Scratch Jr,' a programming language introduced in an interview, provide a low-stakes environment where students can engage with coding concepts without the fear of making mistakes—an especially important factor for EB students who may already experience marginalization due to language barriers. This approach helps students engage meaningfully with programming concepts and nurtures a growth mindset. When students are not bogged down by complex syntax or language barriers, our teacher participants believed that only then can EB students focus on the underlying logic and develop a `feel' for how code works, which is necessary for building a strong foundation and fostering deeper engagement. This aligns with the idea that effective instruction aims to minimize extraneous cognitive load and optimize germane cognitive load to facilitate learning \citep{martin_load_2023, van_merrienboer_cognitive_2010}.


\textbf{The Role of AI:} While not explicitly raised by teacher participants in this study, the potential influence of AI tools (e.g., ChatGPT) on programming pedagogy, particularly for EB students, is an important point for discussion. Observations from outside this study have indicated that EB students are increasingly utilizing AI tools to aid in code comprehension. This raises important questions about the interaction between programming language design and the use of AI in educational settings. It is plausible that the abstraction inherent in BBPLs and other visual programming environments may present challenges for direct interaction with AI code interpretation tools that rely on text-based input. Future research should investigate how diverse programming environments facilitate or impede the integration of AI-assisted learning and the subsequent effects on EB students' development as programmers. This exploration should encompass the potential limitations and the opportunities for leveraging AI to support EB students in programming education, such as AI-driven code translation or explanation tools.

\section{Conclusion}
Teachers play a pivotal role in designing programming instruction that addresses the immediate needs of emergent bilingual (EB) students while preparing them for long-term career readiness. This study highlights the importance of scaffolding through programming languages (PLs) that resemble natural language and visual tools like block-based programming languages (BBPLs), which provide accessible entry points for EB students. However, a tension exists between the need for scaffolding and the goal of equipping students with skills for `real-world' programming careers, which often rely on text-based programming languages (TSPLs). Programming environments must support a gradual transition from BBPLs to TSPLs, enabling students to build confidence while advancing to more rigorous tasks.

Flexibility and multilingual support in programming tools, alongside engagement strategies that emphasize creativity, exploration, and meaningful applications, are critical for maintaining student motivation. Teachers also need resources and training to navigate competing priorities and balance accessibility, rigor, and engagement in their classrooms. By integrating these approaches, teachers can foster equitable yet practical learning experiences that empower EB students to succeed in programming and beyond. Future research should explore how instructional strategies and programming environments can better support these transitions, ensuring all students thrive in a technology-driven world.

    \subsection{Limitations}
    The findings of this study are drawn from a sample of eight K-12 computing teachers in the US who teach computing in classrooms, including EB students. While we interviewed teachers across various K-12 levels, geographic regions (Michigan and Texas), and course types, allowing for a range of perspectives, which likely strengthens the transferability of our findings to a broader teacher population, the sample size restricts the extent to which these findings can be generalized to the broader population of K-12 computing teachers \cite{smith_generalizability_2018}. We also recognize that other teachers may have varying programming proficiency levels,  experience with EB students, and familiarity with other languages their EB students may speak. Furthermore, teachers in our study had varying levels of experience with both EB students and different programming languages, which may have influenced their perspectives and the trade-offs they emphasized.     
    A more diverse and extensive sample, encompassing a wider range of racial/ethnic backgrounds, geographic locations, and teaching contexts, would improve the generalization of these results. Future research with a larger and more diverse sample of teachers is needed to confirm and expand upon these findings, as well as to explore additional factors that may influence teachers' decisions, such as school resources, district policies, and specific student demographics.

\section*{Acknowledgments}
We first thank the teachers for their participation, insights, and support on this work. This material is based upon work supported by the National Science Foundation Graduate Research Fellowship Program under Grant No. DGE-2241144. Any opinions, findings, conclusions, or recommendations expressed in this material are those of the author(s) and do not necessarily reflect the views of the National Science Foundation.

\bibliographystyle{PLATEAU-Reference-Format}
\bibliography{plateau25-refs}

\end{document}